# Privacy-Preserving Deep Learning Model for Covid-19 Disease Detection

Vijay Srinivas Tida
University of Louisiana at Lafayette
vijaysrinivas.tida1@louisiana.edu

Sai Venkatesh Chilukoti
University of Louisiana at Lafayette
saivenkatesh.chilukoti1@louisiana.edu

Sonya Hsu
University of Louisiana at Lafayette
hsiu-yueh.hsu@louisiana.edu

Xiali Hei
University of Louisiana at Lafayette
xiali.hei@louisiana.edu

**Abstract**

*Recent studies demonstrated that usage of X-ray radiography images showed higher accuracy than Polymerase Chain Reaction (PCR) testing for COVID-19 detection. Deep learning models through transfer learning using X-ray radiography images helped increase the speed and accuracy of determining COVID-19 cases compared to the traditional approaches. However, due to Health Insurance Portability and Accountability (HIPAA) compliance, the hospitals were unwilling to share patient data directly with the public due to privacy concerns. In addition, general deep learning models are prone to extract sensitive information from the input data, resulting in vulnerability to the patient's privacy and cannot be used for real-world situations. We propose a privacy-preserving deep learning model for COVID-19 disease detection using a Differential Private (DP) Adam with a transfer learning approach to avoid these problems. Results indicated that the proposed model surpassed the performance of the existing models and obtained 84% of accuracy with a privacy loss ϵ of 10 for classifying the infected patients with COVID-19.*

*Keywords: deep learning, DP Adam optimizer, privacy loss, pre-trained model, transfer learning*

## 1. Introduction

COVID-19 is an infectious disease caused by the SARS-CoV-2 virus that will affect the respiratory system of the infected patient (*World Health Organization COVID-19*, 5/26/2022 accessed). WHO declared the COVID-19 outbreak a global pandemic in March 2020 (Cucinotta et al., 2020). Precise earlier detection can limit the spread of COVID-19, while it is challenging for individuals in developing nations to access robust testing equipment. (Y. Wang et al., 2020) stated that the images of chest x-ray radiography outperformed traditional laboratory testing for identifying COVID patients. Thus, X-ray imaging can be considered the best choice for detecting the presence of COVID-19, and it can help limit the spread of the disease to a great extent considering the global crisis.

Deep learning models showed promising results in detecting the COVID-19 infected patient with the help of chest radiography images. Thus, the deep learning models for COVID-19 detection are selected such that they can be easily deployed to handheld devices like smartphones (Naylor, 2018). Researchers need Chest X-rays radiography (CXR) (Cleverley et al., 2020) as the primary data source for creating a dataset and developing a better model to predict the disease. However, General Data Protection states that the patient data should maintain privacy to provide security from malicious attacks (Voigt et al., 2017; Liu et al., 2020). Therefore, hospitals are unwilling to share the data due to privacy concerns of the patients according to their state laws to avoid data breaches (Landi, 6/14/2022 accessed). Furthermore, deep learning models are also prone to attacks by intruders, especially medical data, which can harm society and cause problems for hospital management if they share the models trained on the available data samples (M. Wu et al., 2020). Therefore, there is a dire need for modifying deep learning models to protect the patient's information from information stealing and also should benefit society in detecting diseases.

Prior research primarily focused on improving the accuracy of detecting COVID-19 disease by assuming the data samples are available publicly (Shoeibi et al., 2020). However, it is not easy to get patients' data samples due to privacy laws enacted in some countries. To solve this problem, some researchers focused on maintaining privacy by injecting noise into the output labels using the Private Aggregation of Teacher Ensembles (PATE) approach (Lange, n.d.; Müftüoğlu et al., 2020). The disadvantage of these previous works is that privacy loss will be more for obtaining the best performing model for COVID-19 detection. In this proposed work, we will try to create a balance between privacy loss and performance using a DP Adam optimizer injected into the neural network. The final goal of the proposed model will be attaining the minimum privacy loss with maximum performance. Popular frameworks like Opacus in PyTorch framework (Yousefpour et al., 2021) and Tensorflow privacy in Tensorflow framework (Abadi et al., 2016) tried to solve these issues by developing differential private deep learning models. These frameworks help inject the noise at the different stages in the deep learning model during the training phase. To the best of our knowledge, we are the first to propose the privacy-preserving model for COVID-19 detection using a DP Adam optimizer and obtain better results than previous deep learning models.

In summary, the contributions of the paper are explained as follows:

- We propose a privacy-preserving deep learning model for COVID-19 detection using chest X-ray images with the help of a DP Adam optimizer. The proposed model will help limit attacks from stealing the patients' information from intruders.

- We finetune the EfficientNet model versions to get the best-performing model for COVID-19 detection without injecting privacy constraints during training. The obtained model is considered the privacy-preserving model for COVID-19 detection.

- We investigate the performance of the proposed model by varying the number of trainable layers with privacy constraints injected into the Adam optimizer during training. Our experimental results show that increasing the number of trainable layers significantly improved accuracy.

The rest of the paper is organized as follows: Section 2 overviews the background and the previous related research works. Then, Section 3 explains the process of designing the privacy-preserving deep learning model for COVID-19 detection. Then, Section 4 presents the results obtained from the proposed model, and Section 5 discusses future research directions for the proposed work. Finally, we conclude the presented work in Section 6.

## 2. Background and literature review

Deep learning models have significantly contributed to medical applications primarily related to image-based diagnosis. The performance of deep learning models for computer vision problems in the medical field is promising (Lundervold et al., 2019). Furthermore, deep learning models showed their importance in classifying diseases such as lung disease, malaria, various cancers like breast and lung cancer, etc (Kieu et al., 2020). Last but not least, privacy-preserving deep learning models became necessary in the medical field because of privacy concerns.

### 2.1. Differential Privacy

Differential Privacy (DP) is considered a standard technique for quantifying the disclosure of individual information from the given input data (Dwork et al., 2006; Dwork, Roth, et al., 2014). The primary purpose of DP is to limit any particular user's information by perturbing the model's parameters during training in the released model. DP can be quantified using two parameters, namely $\epsilon$ and $\delta$. $\epsilon$ measures the difference in change between two datasets that differ by a single sample, and $\delta$ is the probability of information accidentally being leaked. The smaller $\epsilon$ and $\delta$ provide more privacy for the individual. DP can be defined as below: A randomized algorithm A: D $\rightarrow$ R with domain D and range R is ($\epsilon, \delta$)- differential private if for any subset of outputs S $\subseteq$ R and for any two adjacent inputs d, d´$\in$ : $Pr[A(d) \in S] \leq e^{\epsilon} Pr[A(d´) \in S] + \delta$. In a deep learning context, differential privacy can be treated as sharing general information about a dataset by securing individual information. The researcher's main goal is to find the trade-off between the maximum information extraction needed to perform the given task without compromising the individual's privacy (Dwork, 2008). Differential privacy has become critical usage in various areas especially in government agencies (Drechsler, 2021), the health care sector (Ficek et al., 2021), service providers (T. Wang et al., 2020; Gai et al., 2019) etc due to data privacy regulations.

### 2.2. Differential private noise mechanisms

Two popular noise mechanisms can be applied to deep learning models (J. Zhao et al., 2019), namely, Laplacian Mechanism (L. Zhao et al., 2019), and Gaussian Mechanism (McMahan et al., 2018). Gaussian Mechanism (GM) achieves better privacy than the Laplacian mechanism. This is because GM adds noise drawn from normal distribution to the given input data. The Gaussian Mechanism will not satisfy the pure $\epsilon$-differential privacy but does satisfy ($\epsilon$, $\delta$)-differential privacy. Therefore, the Gaussian Mechanism that is $\epsilon$, $\delta$-differential private provided with standard deviation $\sigma$ should satisfy the Equation 1.

$$\sigma \geq \sqrt{2\log(\frac{1.25}{\delta})}\frac{\Delta_2 f}{\epsilon} \qquad (1)$$

In (Bu et al., 2020), the authors released a deep learning model with Differential Privacy that was added using Gaussian noise code in the Tensorflow privacy framework. Also, the GM is popular and has a benefit for training deep learning models using $l_2$ norm instead of $l_1$ norm. Moreover, GM requires lesser noise compared to the Laplacian mechanism. Since the parameters of deep learning models are large in number, GM with $l_2$ sensitivity is used more often than the Laplacian mechanism to get better models.

### 2.3. Rényi Differential Privacy

Rényi Differential Privacy (RDP) is considered as the new generalization of $\epsilon$- differential privacy introduced in (Mironov, 2017) by Marinov that is comparable to the ($\epsilon$, $\delta$) version. Its main advantage is that it is easier to interpret than ($\epsilon$, $\delta$)-DP. It is considered a natural relaxation of differential privacy based on Rényi divergence. Given the random variables X and Y, which can take on n possible values, each with positive probabilities $p_i$ and $q_i$ respectively, then the Rényi divergence of X from Y can be defined by the Equation 2:

$$D_\alpha(X||Y) = \frac{1}{(\alpha-1)} \log\left(\sum_{i=1}^{n} \frac{p_i^\alpha}{q_i^{\alpha-1}}\right) \qquad (2)$$

for $\alpha > 0$ and not equal to 1. However, special cases will be obtained when $\alpha = 0, 1, \infty$. As $\alpha$ converges to 1, $D_\alpha$ converges to the Kullback-Leibler divergence (Joyce, 2011). RDP has nice composition properties that help compute the model's total privacy budget. RDP privacy falls somewhere in between $\epsilon$-DP and ($\epsilon, \delta$)-DP.

### 2.4. Literature review

In (Shoeibi et al., 2020) gave a comprehensive overview of COVID-19 disease detection using various deep learning models. However, some models have achieved 100% accuracy using X-ray, CT-scan, and LUS images with limited data samples of below 5,000. Later, the Kaggle website made a COVID-19 detection challenge and showed that the EfficientNet pre-trained model performed well compared to other models (Website, 5/26/2022 accessedb). (Chetoui et al., 2021) used the EfficientNet model for finetuning and achieved the accuracy of 99% for detecting COVID-19 samples. (Chowdhury et al., 2020) also proposed an ensemble of deep convolutional neural networks based on EfficientNet obtained 97% of accuracy for COVID detection. (Jaiswal et al., 2020) proposed the CovidPen model for detecting COVID-19 disease using Chest X-ray and CT-scan showed the highest accuracy of 96% and 85%, respectively. Also, (Alquzi et al., 2022) finetuned the EfficientNet B-3 model for COVID-19 disease detection and got 99% of test accuracy. (Lee et al., 2022) enhanced the performance of the DenseNet201 for COVID-19 detection using data augmentation, adjusted class weights, and early stopping and finetuning techniques. Results showed the outstanding performance of 99.9% accuracy with a better F1 score of 0.98. (Luz et al., 2022) generated the normal, pneumonia, and COVID-19 infected samples using Deep Convolutional Generative Adversarial Networks and achieved the promising result of 0.95 in terms of area under the curve (AUC) with the help of EfficientNet B3 model. (Huang et al., 2022) obtained the highest performance from the proposed LightEfficientNetV2 model of about 97.73% accuracy. However, designing deep learning models without differential privacy can threaten individuals, and because of government regulations in certain countries, hospitals are unwilling to share the data with the public (M. Wu et al., 2020; R. Wu et al., 2012). So there is a dire need to develop privacy-preserving deep learning models to benefit society without compromising the privacy of the individuals.

The algorithms related to differential privacy are continuously developed, especially in healthcare applications (Ficek et al., 2021). Therefore, hospital management can share the trained differential private models with the public will benefit society. PATE approach was used by (Mu¨ftu¨og˘lu et al., 2020) for privacy-enhanced COVID-19 disease detection and showed the accuracy of 71% with $\epsilon$ of 5.9 and the original model without applying privacy constraints showed an accuracy of 94.7%. (Lange, n.d.) also used the PATE approach for COVID-19 disease detection and showed the accuracy of 75% with $\epsilon$ of 10. (Boulila et al., 2022) proposed the privacy-preserving deep learning model approach by modifying the input images using a homomorphic encryption technique and trained using MobileNetV2. The proposed model showed a comparable accuracy of 93.3% with the model with unencrypted input images. However, most of the existing privacy-preserving models showed limited performance because of problems in extracting the required information due to conditions during the training phase. So, there is necessary to design a deep learning model to address this issue without compromising data privacy.

### 3. Methodology

This section will explain the process of building the privacy-preserving deep learning model for COVID-19

disease detection. First, the procedure of selecting the best pre-trained model for the chosen application will be illustrated. Later in this section will give an overview of the general structure for the privacy-preserving deep learning model to enhance the privacy of the input samples during training using a differential private (DP) optimizer. After the working of DP Adam optimizer is explained in detail with an algorithm. At the end of this section will demonstrate the steps for obtaining the privacy-preserving deep learning for detecting the infected COVID-19 samples.

### 3.1. Pretrained model selection and its overview

Previous works show that the EfficientNet pre-trained model has proven to be the best compared to other deep learning models for COVID-19 disease detection from X-ray radiography images. (Tan et al., 2019) proposed the various versions of EfficientNet models based on scaling the width, depth, and resolution with a fixed ratio. Thus, this property helps the researchers select the best version of the EfficientNet model based on the chosen application. Additionally, the EfficientNet model requires fewer parameters than the existing pre-trained models because of having depthwise separable and pointwise convolution layers instead of conventional convolution layers. This replacement can help the EfficientNet model to deploy on mobile devices, which can be helpful for healthcare applications.

Furthermore, the addition of squeeze and excitation blocks helped the model improve performance by mapping the channel dependency across the input feature maps with access to global information (Hu et al., 2018). There are multiple versions of EfficientNet models, namely from B0 to B7. Neural Architecture Search initially defines the structure for the EfficientNet B0 model. Later, with the help of the compound scaling technique and the information from the previous model versions was used to get the subsequent versions.

### 3.2. Privacy-preserving deep learning model

The general structure of the chosen privacy-preserving deep learning model having a noisy optimizer is depicted clearly in Fig. 1. In this process, the noisy optimizer will replace the general optimizer, and all other steps for training the neural network will still hold the same. During the training phase of the model, the preprocessed image samples are fed directly into the model. During the preprocessing, the images are converted to the corresponding size accordingly to the model requirement. After the forward propagation process, the model predicts the output for the input image samples. Then, the loss function gives the score based on the actual and predicted values for the given inputs from the dataset and the model respectively. Later, the optimizer updates the network weights by calculating the gradients corresponding to the selected loss function (Sun et al., 2019). The obtained gradients act as information extractors from the given image samples. Thus, clipping the gradients and the injection of the noise selectively into the optimizer limits the information extraction, and this process converts the general optimizer into a noisy optimizer.

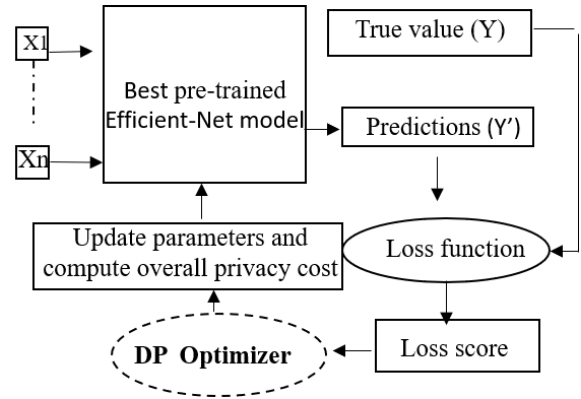

**Figure 1. The general structure of privacy-preserving deep learning model**

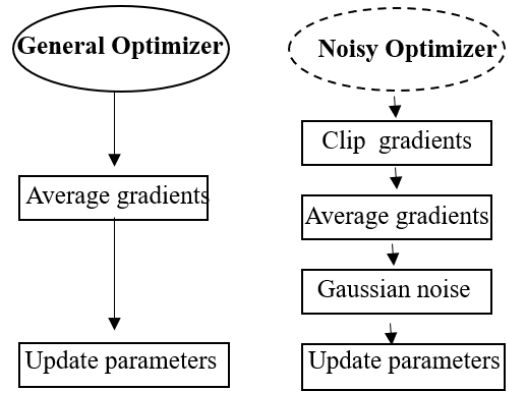

**Figure 2. Overview of a general optimizer and a noisy optimizer**

Fig. 2 shows the difference between the general optimizer and the noisy optimizer. The gradients from the individual samples in a batch are averaged directly before updating the parameters for the next iteration using a general optimizer. This process causes the information extraction from the specific samples can be more which can cause harm to the related subjects by the intruders if the trained models are available publicly. To avoid this situation, adding noise at the different stages in the neural network model can limit the leakage of sensitive information from the given input samples. Among them, adding the noise at the optimizer level will help reduce the overfitting during the training phase of the model, resulting in improved performance of the trained model. Additionally, limiting the gradient information flow at this level help to achieve maximum performance without compromising the privacy of the given input data. Thus a noisy optimizer is made by clipping the gradients before averaging and adding the noise while updating the parameters. The optimizer selection also plays a crucial role in faster convergence of the deep neural network model. Adam optimizer showed many advantages like efficient computation, requiring less memory, and time for hyperparameter tuning is considered for the proposed model. Computing the overall privacy cost at the end makes the noisy optimizer to DP optimizer.

### 3.3. DP Adam optimizer

---

**Algorithm 1** DP Adam Optimizer

**Input:** dataset $S = x_1, ..., x_n$, loss function $l(\theta, x)$
**Parameters:** initial weights $\theta_0$, learning rate $\eta_t$, sub-sampling probability $p$, number of iterations $T$, noise scale $\sigma$, gradient norm bound $R$, momentum parameters $(\beta_1, \beta_2)$, initial momentum $m_0$, initial past squared gradient $u_0$, and a small constant $\epsilon > 0$.

**for** $t = 0, \ldots, T-1$ **do**

    Take a Poisson subsample $I_t \subseteq \{1, \ldots, n\}$ with subsample probability $p$

    **for** $i \in I_t$ **do**

        $v_t^{(i)} \leftarrow \nabla_\theta l(\theta_t, x_i)$
        $\bar{v}_t^{(i)} \leftarrow v_t^{(i)} / \max\{1, \|v_t^{(i)}\|_2 / R\}$
        ▷ — Clip gradient

    $\tilde{v}_t \leftarrow \frac{1}{|I_t|}\left(\sum_{i \in I_t} \bar{v}_t^{(i)} + \sigma R \cdot \mathcal{N}(0, I)\right)$
    ▷ —Apply Gaussian Mechanism

    $m_t \leftarrow \beta_1 m_{t-1} + (1 - \beta_1)\tilde{v}_t$
    $u_t \leftarrow \beta_2 u_{t-1} + (1 - \beta_2)(\tilde{v}_t \odot \tilde{v}_t)$
    ▷ – $\odot$ is the Hadamard product

    $w \leftarrow m_t / (\sqrt{u_t} + \epsilon)$ ▷ – Component wise division
    $\theta_{t+1} \leftarrow \theta_t - \eta_t w_t$

**Output:** $\theta_t$ and compute the overall privacy cost $(\epsilon, \delta)$ using a privacy accountant mechnaism

---

The general implementation of the DP Adam optimizer is used to train the privacy-preserving deep learning model in Algorithm 1 (Bu et al., 2020). This optimizer follows similar steps to the general Adam optimizer with minor changes. For this algorithm, input will be the samples $S$ from the dataset and the selected loss function $l(\theta, x)$ based on the application. The optimizer will update the neural network model parameters considered as output for every iteration of given batch samples. This updating process will have two additional steps for the noisy Adam optimizer compared to the general procedure. First, we select Poisson subsamples from the given data with a probability of p. Then, after choosing the specific sample, its gradient is calculated, and the obtained value is clipped for all the selected batch samples. Next, noise is added after getting the clipped gradients for each sample using the Gaussian Mechanism. Then after the first momentum and second momentum, estimates ($m_t$ and $u_t$) are calculated using momentum parameters $\beta_1$ and $\beta_2$, respectively. In the end, parameters are updated using component-wise division and can be used for the next iteration. But before starting the next iteration, the optimizer must compute the privacy cost based on the given privacy accountant mechanism. This step will convert the noisy Adam optimizer to the DP Adam optimizer.

### 3.4. Privacy-Preserving model for predicting COVID-19

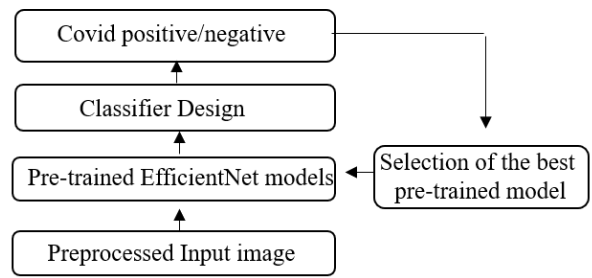

**Figure 3.** The selection process of the best version among pretrained Efficient Net models for COVID-19 disease detection

For obtaining the privacy-preserving model for COVID-19 disease detection, there is a need to get the best-performing deep learning model without privacy constraints injected into it. The general procedure for identifying the best performing model can be seen in Fig. 3. Here, images are preprocessed according to the requirement of the selected version and fed as input to the model. Then classifier design is made by placing the fully connected layer with one output neuron according to the chosen model for COVID-19 disease prediction. The hyperparameters are obtained from the best-performing model in the Kaggle competition for COVID-19 disease detection using X-ray radiography images (Sharman, 5/26/2022 accessed). The finetuning process is made using the above hyperparameters for all EfficientNet models to achieve the best-performing model.

EfficientNet-B2 showed a higher performance based on test accuracy for predicting the infected samples and is used as the base structure for designing a privacy-preserving deep learning model. First, the privacy engine is attached to the Adam Optimizer directly with the help of the Opacus framework. Through the privacy accounting mechanism, this framework will keep track of the privacy budget during the training stage based on Re´nyi Differential Privacy in the background. The privacy engine helps identify the privacy loss spent at any given point by adding early stopping and real-time monitoring features. The model validation module from the Opacus library makes the given deep neural network compatible with the privacy engine attached to the optimizer. These changes include changing the particular layers like Batch Normalization (Bjorck et al., 2018) that have information from all the samples with Instance Normalization (Ulyanov et al., 2016) and Group Normalization (Y. Wu et al., 2018) to keep track of individual sample gradients during the training stage. Then the proposed model is finetuned by varying the trainable layers to find the best-performing model with the above characteristics. The obtained results showed a significant improvement in the performance with an increase in the number of trainable layers.

Healthcare services can use the above process by training the model without adding privacy constraints to achieve the maximum performance for the given task as an initial step. Then later, the privacy-preserving model can be developed with the same structure obtained from the previous step by adding the privacy constraints into the model. Finally, the privacy-preserving is finetuned accordingly to achieve comparable performance with the original model. After, the obtained privacy-preserving model can be shared with the public by discarding the original model. This process can help low-income countries rely on deep learning models for disease detection without needing to access robust testing equipment.

## 4. Results

### 4.1. Dataset description

For evaluating the proposed COVID-19 disease detection model, we used a dataset from the Kaggle website (Website, 5/26/2022 accesseda). The samples are X-ray radiography images from the healthcare centers, available on Kaggle website for researchers.

| Efficient Net Pre-trained model versions | Valid Acc | Test Acc |
|---|---|---|
| B 0 | 0.98 | 0.91 |
| B 1 | 0.98 | 0.94 |
| B 2 | 0.98 | 0.95 |
| B 3 | 0.98 | 0.94 |
| B 4 | 0.98 | 0.94 |
| B 5 | 0.98 | 0.91 |
| B-6 | 0.98 | 0.90 |

**Table 1. Validation Accuracy (Valid Acc) and Test Accuracy (Test Acc) for different versions of Efficient-Net Model**

The total number of samples in the dataset used for training is 30,492; among them, 16,490 are COVID-positive, and 13,992 are COVID-negative. They provided only 400 samples to evaluate the model's performance on test data. However, 20% samples from the training dataset were used for validation purposes to follow the standard testing process. The model designed is a supervised model. The samples are labeled 1 or 0 based on COVID-infected samples or not. Since the dataset is only slightly imbalanced, we didn't use any data augmentation techniques to increase the number of samples to train the model.

### 4.2. Evaluation of pretrained EfficientNet model versions

We divided the data samples into 80% for training and 20% for validation from the selected dataset. The number of data samples used for testing is 400, which were given separately in the dataset directly for evaluating the designed model. Therefore, the accuracy obtained from the validation dataset can be considered validation accuracy, and the accuracy obtained from test data samples is deemed as test accuracy. We used Binary Cross-Entropy Loss as the loss criterion and Adam optimizer for finetuning the pre-trained EfficientNet model versions. The learning rate of the optimizer is 0.08, the number of epochs is 30, and the batch size of 32 is used for training the model.

Table 1 represents the validation and test accuracy obtained for different versions of the EfficientNet model. The EfficientNet model versions are trained generally without injecting privacy constraints using the PyTorch framework (Paszke et al., 2019). Since the validation accuracy for all the models remains the same for all versions with 98% accuracy, we considered test accuracy as the best choice for selecting the best-performing model for further evaluation. Among all the model versions, EfficientNet-B2 showed the highest performance for test data with 95% accuracy and is considered the base structure for the privacy-preserving model. Since all EfficientNet model versions performed well on validation accuracy, any model version can be used for developing the privacy-preserving deep learning model. For simplicity in choosing one model, we rely on the test accuracy, but it is not required to select the best-performing model.

### 4.3. Evaluation of the privacy-preserving deep learning model

| $\epsilon$ | Max grad norm | 1st case | | 2nd case | |
|---|---|---|---|---|---|
| | | Valid Acc | Test Acc | Valid Acc | Test Acc |
| 1.0 | 1 | 0.93 | 0.80 | 0.91 | 0.74 |
| 2.0 | 1 | 0.93 | 0.83 | 0.92 | 0.81 |
| 10.0 | 1 | 0.94 | 0.82 | 0.93 | 0.80 |
| 100 | 1 | 0.94 | 0.83 | 0.95 | 0.86 |
| 1000 | 1 | 0.93 | 0.84 | 0.95 | 0.89 |
| 1.0 | 0.8 | 0.93 | 0.82 | 0.91 | 0.76 |
| 2.0 | 0.8 | 0.93 | 0.82 | 0.93 | 0.79 |
| 10 | 0.8 | 0.92 | 0.83 | 0.95 | 0.84 |
| 100 | 0.8 | 0.94 | 0.83 | 0.95 | 0.86 |
| 100 | 0.8 | 0.94 | 0.83 | 0.95 | 0.89 |
| 1000 | 0.8 | 0.94 | 0.83 | 0.90 | 0.70 |
| 1.0 | 0.6 | 0.93 | 0.80 | 0.93 | 0.79 |
| 2.0 | 0.6 | 0.93 | 0.80 | 0.94 | 0.84 |
| 10 | 0.6 | 0.93 | 0.80 | 0.94 | 0.83 |
| 100 | 0.6 | 0.93 | 0.83 | 0.95 | 0.90 |
| 1000 | 0.6 | 0.94 | 0.83 | 0.92 | 0.78 |
| 1.0 | 0.4 | 0.92 | 0.78 | 0.93 | 0.78 |
| 2.0 | 0.4 | 0.93 | 0.80 | 0.93 | 0.84 |
| 10 | 0.4 | 0.93 | 0.81 | 0.94 | 0.85 |
| 100 | 0.4 | 0.93 | 0.82 | 0.95 | 0.88 |
| 1000 | 0.4 | 0.93 | 0.84 | 0.93 | 0.84 |

**Table 2.** Validation Accuracy (Valid Acc) and Test Accuracy (Test Acc) for (i) by finetuning the last block of EfficientNet-B2 model (1st case) (ii) by finetuning the last two blocks of EfficientNet-B2 model (2nd case)

| Authors | Method | Epsilon | Accuracy |
|---|---|---|---|
| (Müftüoğlu et al., 2020) | PATE | 5.9 | 71% |
| Proposed | DPAdam | 6 | 82% |
| (Lange, n.d.) | PATE | 10 | 75% |
| Proposed | DPAdam | 10 | 84% |

**Table 3.** Comparison of Test Accuracy for Differential Privacy applied to the Deep Learning models

For evaluating the privacy-preserving deep learning model, we finetuned the EfficientNet-B2 model, which was considered the best version from the previous analysis. The same data distribution, along with the loss function, is still carried out here to evaluate the model's performance. The private engine attachment to the optimizer version will change to the DP Adam optimizer with the constraints explained in Section 3.3 during the training phase. The only hyperparameter added for the proposed model is δ, which is considered the probability of failure, and its value is 1e$^{-5}$, while all others remain the same for training the model.

The privacy loss $\epsilon$ of zero means that the model is considered random with utmost privacy, whereas the privacy budget $\epsilon$ is inf, indicating no privacy for the trained model. Maximum gradient normalization is the maximum amount of information obtained from a given sample through gradients during the training process. This parameter can be regarded as clipping the gradients to the given constant value when it is out of the given range to limit the information extraction. As a researcher, we need to find the limit of $\epsilon$ such that the data privacy leakage should be minimum while extracting the maximum information from the given input data for training the privacy-preserving model. Opacus library provides the modulator fix option to make the normal model suitable for the privacy-preserving application.

Initially, the fully connected layer was made trainable, resulting in a limited performance with 60-65% of accuracy from the test data samples. Later the last block of the selected model is made trainable along with the fully connected layer, which can be seen in Table 2. Results showed that validation and test accuracy increase when $\epsilon$ increases for fixed maximum gradient normalization values. The proposed model achieved the highest test accuracy of 84%, but the privacy loss $\epsilon$ needed is 1,000. Similarly in the second case, the last two blocks of the selected model with a fully connected layer are made trainable and noted down the corresponding performance. Compared to the previous version, this model showed a higher accuracy of 84% with a privacy loss $\epsilon$ of 10. The highest test accuracy obtained by varying the last two blocks is 90%, with a privacy loss $\epsilon$ of 100 and a maximum gradient normalization value of 0.6. An interesting phenomenon happened in the second case: when maximum gradient normalization varied, the accuracy sometimes decreased even if privacy loss $\epsilon$ increased. This can be observed for privacy loss $\epsilon$ changed from 100 to 1000 for maximum gradient normalization falls below 1. The reason might be increasing the number of trainable layers might need more gradient information, even though allowing more privacy loss to obtain higher accuracy (Andrew et al., 2021). The results show that the privacy-preserving deep learning model achieves comparable performance to the model without injecting any privacy constraints.

Table 3 compares the accuracy obtained for COVID-19 detection when privacy constraints are injected into the model. The previous approach proposed by (Müftüoğlu et al., 2020) showed an accuracy of 71% using PATE analysis for COVID-19 detection. The proposed model showed a higher accuracy of 82% with the privacy loss nearly to 6. Also, (Lange, n.d.) showed an accuracy of 75% while the proposed approach showed 84% accuracy with $\epsilon$ of 10. Our proposed method showed the best performance in both cases compared to these works with the same privacy loss. PATE analysis needs accessibility to the public data samples for training the model using semi-supervised settings. However, in real-life scenarios, the accessibility of data samples from health care centers might not be readily available, and applying PATE is impossible for these situations (Uniyal et al., 2021). In the case of the proposed privacy-preserving model, there is no need for publicly available samples to train the model. The healthcare centers can train the models with privacy constraints, and the user can use their model directly without the need to collect the samples for training purposes.

## 5. Future work

The healthcare applications like cancer detection, diabetic retinopathy grading analysis, etc. (Sai Venkatesh et al., 2022; Chilukoti et al., 2022; Islam et al., 2022) can use the proposed privacy-preserving deep learning model with a DP Adam optimizer to achieve better performance. Furthermore, the proposed work can also be applied in federated learning settings by making individual privacy-preserving models before aggregating them into the central server (Yang et al., 2019; Wei et al., 2020). In the near future, secure multiparty computation (Cramer et al., 2015) can also be integrated to train the deep learning models by encryption from Crypten library (Knott et al., 2021) proposed by Facebook. Moreover, the fixed clipping method in the proposed model can be replaced with the adaptive clipping mechanism and adding the Gaussian noise before aggregating can improve the performance (Andrew et al., 2021). To further enhance the performance of the privacy-preserving model, honest hyperparameter selection is essential and can be achieved through the approach provided by (Mohapatra et al., 2022). The proposed-privacy preserving model can also be applied to Natural Language Processing (NLP) applications like Spam Detection, Fake-news Detection (Tida & Hsu, 2022; Tida et al., 2022) etc.

## 6. Conclusion

individual privacy has become crucial to be compliant the government regulations and HIPPA. There is high demand for privacy-preserving deep learning models by limiting the extraction of information. This manuscript provided a DP Adam optimizer based privacy-preserving deep learning model for detecting COVID-19 samples using X-ray radiography images. The proposed model achieved 84% of accuracy with privacy loss $\epsilon$ of 10 by finetuning the last two blocks of the EfficientNet-B2 model and achieved better performance than the previous works. Furthermore, the results showed that increasing the number of trainable layers improved the performance of the proposed model with the same privacy loss $\epsilon$. The proposed approach serve as the benchmark for other health care applications, and contribute other researchers in getting the best-performing model with the help of a DP Adam optimizer.

## Acknowledgement

We sincerely thank Drs. Md Imran Hossen and Liqun Shan for their valuable time for the suggestions and for his dedicated help in organizing the manuscript effectively. This work is supported in part by the US NSF under grants OIA-1946231 and CNS-2117785.

## References

Abadi, M., et al. (2016). Deep learning with differential privacy. In *Proceedings of the 2016 acm sigsac conference on computer and communications security* (pp. 308–318).

Alquzi, S., et al. (2022). Detection of covid-19 using efficientnet-b3 cnn and chest computed tomography images. In *International conference on innovative computing and communications* (pp. 365–373).

Andrew, G., et al. (2021). Differentially private learning with adaptive clipping. *Advances in Neural Information Processing Systems*, *34*, 17455–17466.

Bjorck, N., et al. (2018). Understanding batch normalization. *Advances in neural information processing systems*, *31*.

Boulila, W., et al. (2022). Securing the classification of covid-19 in chest x-ray images: A privacy-preserving deep learning approach. *arXiv preprint arXiv:2203.07728*.

Bu, Z., et al. (2020). Deep learning with gaussian differential privacy. *Harvard data science review*, *2020*(23).

Chetoui, M., et al. (2021). Explainable covid-19 detection on chest x-rays using an end-to-end deep convolutional neural network architecture. *Big Data and Cognitive Computing*, *5*(4), 73.

Chilukoti, et al. (2022). Diabetic retinopathy detection using transfer learning from pre-trained convolutional neural network models.

Chowdhury, N. K., et al. (2020). Ecovnet: An ensemble of deep convolutional neural networks based on efficientnet to detect covid-19 from chest x-rays. *arXiv preprint arXiv:2009.11850*.

Cleverley, J., et al. (2020). The role of chest radiography in confirming covid-19 pneumonia. *bmj*, *370*.

Cramer, R., et al. (2015). *Secure multiparty computation*. Cambridge University Press.

Cucinotta, D., et al. (2020). Who declares covid-19 a pandemic. *Acta Bio Medica: Atenei Parmensis*, *91*(1), 157.

Drechsler, J. (2021). Differential privacy for government agencies–are we there yet? *arXiv preprint arXiv:2102.08847*.

Dwork, C. (2008). Differential privacy: A survey of results. In *International conference on theory and applications of models of computation* (pp. 1–19).

Dwork, C., et al. (2006). Calibrating noise to sensitivity in private data analysis. In *Theory of cryptography conference* (pp. 265–284).

Dwork, C., Roth, A., et al. (2014). The algorithmic foundations of differential privacy. *Found. Trends Theor. Comput. Sci.*, *9*(3-4), 211–407.

Ficek, J., et al. (2021). Differential privacy in health research: A scoping review. *Journal of the American Medical Informatics Association*, *28*(10), 2269–2276.

Gai, K., et al. (2019). Differential privacy-based blockchain for industrial internet-of-things. *IEEE Transactions on Industrial Informatics*, *16*(6), 4156–4165.

Hu, J., et al. (2018). Squeeze-and-excitation networks. In *Proceedings of the ieee conference on computer vision and pattern recognition* (pp. 7132–7141).

Huang, M.-L., et al. (2022). A lightweight cnn-based network on covid-19 detection using x-ray and ct images. *Computers in Biology and Medicine*, 105604.

Islam, M. M., et al. (2022). Differential private deep learning models for analyzing breast cancer omics data. *Frontiers in oncology*, *12*.

Jaiswal, A. K., et al. (2020). Covidpen: A novel covid-19 detection model using chest x-rays and ct scans. *Medrxiv*.

Joyce, J. M. (2011). Kullback-leibler divergence. In *International encyclopedia of statistical science* (pp. 720–722). Springer.

Kieu, S. T. H., et al. (2020). A survey of deep learning for lung disease detection on medical images: state-of-the-art, taxonomy, issues and future directions. *Journal of imaging*, *6*(12), 131.

Knott, B., et al. (2021). Crypten: Secure multi-party computation meets machine learning. *Advances in Neural Information Processing Systems*, *34*.

Landi, H. (6/14/2022 accessed). *healthcare-data-breaches-hit-all-time-high-2021-impacting-45m-people.* https://www.fiercehealthcare.com/health-tech/healthcare-data-breaches-hit-all-time-high-2021-impacting-45m-people.

Lange, L. (n.d.). Privacy-preserving detection of covid-19 in x-ray images.

Lee, C. P., et al. (2022). Covid-19 diagnosis on chest radiographs with enhanced deep neural networks. *Diagnostics*, *12*(8), 1828.

Liu, X., et al. (2020). Privacy and security issues in deep learning: A survey. *IEEE Access*, *9*, 4566–4593.

Lundervold, A. S., et al. (2019). An overview of deep learning in medical imaging focusing on mri. *Zeitschrift für Medizinische Physik*, *29*(2), 102–127.

Luz, E., Silva, P., Silva, R., Silva, L., Guimarães, J., Miozzo, G., . . . Menotti, D. (2022). Towards an effective and efficient deep learning model for covid-19 patterns detection in x-ray images. *Research on Biomedical Engineering*, *38*(1), 149–162.

McMahan, H. B., et al. (2018). A general approach to adding differential privacy to iterative training procedures. *arXiv preprint arXiv:1812.06210*.

Mironov, I. (2017). Rényi differential privacy. In *2017 ieee 30th computer security foundations symposium (csf)* (pp. 263–275).

Mohapatra, S., et al. (2022). The role of adaptive optimizers for honest private hyperparameter selection. In *Proceedings of the aaai conference on artificial intelligence* (Vol. 36, pp. 7806–7813).

Müftüoğlu, Z., et al. (2020). Differential privacy practice on diagnosis of covid-19 radiology imaging using efficientnet. In *2020 international conference on innovations in intelligent systems and applications (inista)* (pp. 1–6).

Naylor, C. D. (2018). On the prospects for a (deep) learning health care system. *Jama*, *320*(11), 1099–1100.

Paszke, A., et al. (2019). Pytorch: An imperative style, high-performance deep learning library. *Advances in neural information processing systems*, *32*.

Sai Venkatesh, C., et al. (2022). Modified resnetmodel for msi and mss classification of gastrointestinal cancer. In *High performance computing and networking* (pp. 273–282).

Sharman, B. S. (5/26/2022 accessed). *High-accuracy covid 19 prediction from chest x-ray images using pre-trained convolutional neural networks in pytorch.* https://towardsdatascience.com/high-accuracy-covid-19-prediction-from-chest-x-ray-images-using-pre-trained-convolutional-neural-2ec96484ce0.

Shoeibi, A., et al. (2020). Automated detection and forecasting of covid-19 using deep learning techniques: A review. *arXiv preprint arXiv:2007.10785*.

Sun, S., et al. (2019). A survey of optimization methods from a machine learning perspective. *IEEE transactions on cybernetics*, *50*(8), 3668–3681.

Tan, M., et al. (2019). Efficientnet: Rethinking model scaling for convolutional neural networks. In *International conference on machine learning* (pp. 6105–6114).

Tida, V. S., & Hsu, S. (2022). Universal spam detection using transfer learning of bert model. *arXiv preprint arXiv:2202.03480*.

Tida, V. S., et al. (2022). Unified fake news detection using transfer learning of bidirectional encoder representation from transformers model. *arXiv preprint arXiv:2202.01907*.

Ulyanov, D., et al. (2016). Instance normalization: The missing ingredient for fast stylization. *arXiv preprint arXiv:1607.08022*.

Uniyal, A., et al. (2021). Dp-sgd vs pate: Which has less disparate impact on model accuracy? *arXiv preprint arXiv:2106.12576*.

Voigt, P., et al. (2017). The eu general data protection regulation (gdpr). *A Practical Guide, 1st Ed., Cham: Springer International Publishing*, *10*(3152676), 10–5555.

Wang, T., et al. (2020). Edge-based differential privacy computing for sensor–cloud systems. *Journal of Parallel and Distributed computing*, *136*, 75–85.

Wang, Y., et al. (2020). Temporal changes of ct findings in 90 patients with covid-19 pneumonia: a longitudinal study. *Radiology*, *296*(2), E55–E64.

Website, K. (5/26/2022 accesseda). *Chest x-ray images for the detection of covid-19.* https://www.kaggle.com/datasets/andyczhao/covidx-cxr2.

Website, K. (5/26/2022 accessedb). *Covid-19 x-ray image classification.* https://www.kaggle.com/competitions/stat946winter2021/leaderboard.

Wei, K., et al. (2020). Federated learning with differential privacy: Algorithms and performance analysis. *IEEE Transactions on Information Forensics and Security*, *15*, 3454–3469.

*World health organization covid-19.* (5/26/2022 accessed). https://www.who.int/health-topics/coronavirus#tab=tab1.

Wu, M., et al. (2020). Evaluation of inference attack models for deep learning on medical data. *arXiv preprint arXiv:2011.00177*.

Wu, R., et al. (2012). Towards hipaa-compliant healthcare systems. In *Proceedings of the 2nd acm sighit international health informatics symposium* (pp. 593–602).

Wu, Y., et al. (2018). Group normalization. In *Proceedings of the european conference on computer vision (eccv)* (pp. 3–19).

Yang, Q., et al. (2019). Federated learning. *Synthesis Lectures on Artificial Intelligence and Machine Learning*, *13*(3), 1–207.

Yousefpour, A., et al. (2021). Opacus: User-friendly differential privacy library in pytorch. *arXiv preprint arXiv:2109.12298*.

Zhao, J., et al. (2019). Differential privacy preservation in deep learning: Challenges, opportunities and solutions. *IEEE Access*, *7*, 48901–48911.

Zhao, L., et al. (2019). Privacy-preserving collaborative deep learning with unreliable participants. *IEEE Transactions on Information Forensics and Security*, *15*, 1486–1500.